# Minimal Surfaces & the Intermediate State of Type-I Superconductors

Pavol Valko[1], Thomas A. Girard[2]

[1]*Department of Physics, Slovak University of Technology, Ilkovičova 3, 841 04 Bratislava, Slovak Republic*

[2]*Centro de Fisica Nuclear, Universidade de Lisboa, Av. Prof. Gama Pinto 2, 1649-003 Lisboa, Portugal*

The geometry of the intermediate state (IS) flux structures of type-I superconductors, observed during early stages of magnetic flux penetration induced by an externally applied perpendicular magnetic field, is approached in terms of minimal surfaces. This approach follows from the assumption that a local free energy minimization constraint, for each individual flux structure, dominates for sufficiently spatially separated magnetic flux regions, i.e. when over global free energy minimization predicts very low periodicity of sample intermediate state.

## I. INTRODUCTION

The transition between superconducting and normal states in the presence of a magnetic field evolves through a transitional state. Properties of this state depend on numerous parameters and often even on the sample field history [1]. The principal parameter describing superconductive behaviour in a magnetic field is the Ginzburg-Landau parameter ($\kappa$) which is commonly used to classify superconducting materials as type-I or type-II superconductors. For type-I ($\kappa < 1/\sqrt{2}$) macroscopic planar samples (i.e. samples with sufficiently large aspect ratio), an intermediate state (IS) with quasi-periodic pattern of normal and superconducting regions is formed, while type-II ($\kappa > 1/\sqrt{2}$) samples form a mixed state comprising of periodic Abrikosov lattice of microscopic magnetic vortices, each carrying one elementary flux quanta.

The observed type-I structures are rich in geometry: simple flux tubes, complex corrugated patterns, combinations of both or even simple periodic stripes, have been observed [2]. The basic properties of a lamina IS have been described by Landau [3,4], resulting from a global free energy minimization over whole sample volume. Alternatively, Goren and Tinkham [5] considered the IS to consist of a hexagonal array of flux tubes, resulting in about the same free energy as the laminar structure. Another approach describes the IS as current loops [6,7] which flow along the normal (N) – superconducting (S) interfaces, which permits the self-interaction of the currents within a domain as well as in free space. This model reproduces the basic results of the Landau theory for the equilibrium period of straight lamina, and allows the study of instabilities of a flux tubes. More recently, this description was enlarged to include the screening currents in the vicinity of the top and bottom surfaces of the samples [8] and was demonstrated, that the stripes and corrugated patterns are associated with planar samples, and that the flux tube structure is the truly equilibrium, minimum energy structure [9].

In all listed approaches, the observed periodicity is a direct consequence of the global free energy minimization. The contributions of the lost condensation and magnetic energies due to the "flaring out" of the tubes near the sample surfaces can only be calculated for a few simple geometries however, and the individual IS patterns generally remain without a clear explanation.

Rather than elaborating further approach of global free energy minimization, we consider here the geometry of individual flux structures as arising from an additional local free energy minimization constraint. The major geometry-dependent contribution to the free energy, particularly at low applied magnetic field, is the S-N interface energy. The local free energy minimization thus represents the classical problem of interface (surface) area minimization for a given linear contour, constrained by the criterion of flux conservation within each flux bundle individually. The compliance with the dipole character of magnetic field severely reduces the number of available geometries, since only the so-called embedded (no self-intersecting) surfaces are consistent with related Maxwell's equation ($div\ \boldsymbol{B}=0$). We elaborate on this and its consequences in terms of minimal surfaces in Sec- II, and relate the idea to experiment in Sec. III. The results are discussed in Sec. IV, and conclusions formed in Sec. V.

## II. THEORETICAL CONSIDERATIONS

The conventional method to theoretically describe IS structure is via minimization of the free energy of the system globally, i.e. over whole sample volume.

### A  Free Energy

The free energy of a sample in an IS consisting of $n$ identical flux bundles can be expressed as

$$F = -\frac{H_c^2}{8\pi}\left(L_x L_y - nS_B\right)d + \frac{H_c^2}{8\pi}nS_B d + \\ + \frac{H_c^2}{8\pi}nS_{NS}\delta + n\left(F_{cond} + F_{mag}\right) + n^2 F_{int} \quad,(1)$$

where $L_x$, $L_y$ are the sample linear dimensions, $d$ is the sample thickness, $S_B$ is the flux bundle cross section, $H_c$ is the thermodynamic critical magnetic field, and $S_{SN}$ is the S-N interface area (surface); $\delta \sim \xi - \lambda$ is the S-N interface parameter, with $\xi$ ($\lambda$) the material coherence (penetration) lengths.

The first term represents the condensation energy of the superconducting volume fraction, the second is the magnetic energy in the normal zones (bundles) and the third is the S-N interface wall energy. The terms $F_{cond}$ and $F_{mag}$ corresponds to the lost condensation energy and magnetic energy due to opening out of bundles near the top and bottom sample surface. The last term $F_{int}$ is the bundle-bundle interaction energy. When $n$ is small (and the bundle separation large), the interaction term $F_{int}$ can be neglected. The condensation energy term cannot be neglected, but effectively defines only the overall free energy "base-line". The contributions of the remaining terms are proportional to the number of flux structures.

The major contribution to the free energy is therefore given by the geometrical form of the flux structure. The terms $F_{cond}$ and $F_{mag}$ can be calculated only for a few simple geometries, such as laminar, resulting in complicated, S/N - dependent functions [3].

### B. Minimal surfaces

A system which attempts to minimize its free energy by adjusting its geometrical form, is commonly observed in soap foam films bounded by a wire frame. This is a direct consequence of positive surface tensions in such systems. The geometrical shape determination of considerably-separated individual flux bundles is therefore an analogy of the classical Plateau problem, which determines the surface geometry of soap films and other interfaces with a positive surface free energy density [10].

In ideally pure and homogenous type-I superconductor the shape of the "wire frame", from which the superconductor-normal metal interface originates, isn't pre-determined and is determined only by the intersection of the surface representing a minimal energy configuration with the top and bottom sample planes. Surfaces, which are known to possess a minimal surface area for given conditions (such as the loop from which foam film originates), are known as minimal surfaces (MS).

When $V^{n-1} \subset R^n$ is a smooth hyper-surface, within a $M^n$ Riemann manifold, given by the function $x_n = f(x_1,\ldots,x_{n-1})$ defined over a boundary $D \subset R^{n-1}$, the volume functional defined on a set of $f(x_1,\ldots,x_{n-1})$ functions is given as

$$vol(f) = \int_D \sqrt{1+\sum_{i=1}^{n-1} f_{xi}^2}\, dx_1 \wedge \ldots \wedge dx_n. \quad (2)$$

The extreme hyper-surface $V^{n-1}$ for $vol(f)$ can be found via the Euler-Lagrange equation

$$\sum_{i=1}^{n-1} \frac{\partial}{\partial x_i}\left( f_{xi}\left(1+\sum_{j=1}^{n-1} f_{xj}^2\right)^{-\frac{1}{2}}\right) = 0, \quad (3)$$

solutions of which are known as locally MS. In $R^3$, the two dimensional surface $z = f(x,y)$ "volume" functional is

$$A[f] = \iint \sqrt{1+f_x^2+f_y^2}\, dx\, dy, \quad (4)$$

and the corresponding Euler-Lagrange equation reduces to

$$(1+f_x^2)f_{yy} - 2f_x f_y f_{xy} + (1+f_y^2)f_{xx} = 0. \quad (5)$$

There are a number of known solutions of Eq. (5), habitually named after their founders (Enneper's, Henneberg's, Hoffman's, Oliveira's, Costa MS, etc.) or according to their characteristic shape (helicoid, catenoid, gyroid). The simplest MS is a trivial flat surface, while the best known nontrivial MS is the catenoid.

### III. MODEL and ANALOGUES

Consider a situation when a perpendicularly applied field ($H_a$) is raised slightly above the so-called penetration field ($H_p$) of a zero field cooled thin rectangular sample. As seen in Fig. 1, two distinctly different flux-occupied regions are commonly observed in a very pure Type-I superconductors at applied magnetic field in the vicinity of $H_p$.

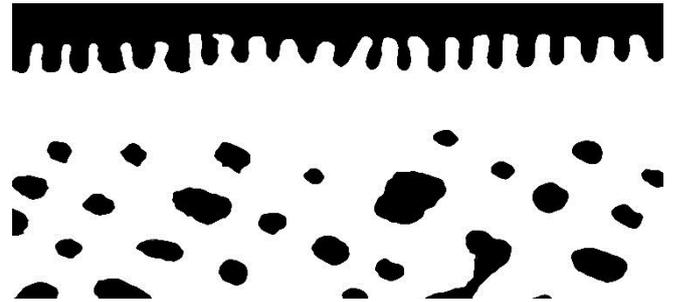

FIG. 1: Sketch of typical flux structures observed in very pure mercury. Motivated by Fig. 8 of Ref. [11].

Curtain like structures are formed near the sample edges, while regions near the sample central line possess typically a tube form. Edge patterns begin their formation already at fields significantly less than $H_p$, when flux lines start to first cut through the sharp upper and lower sample edges. At higher applied fields, flux regions near those surfaces become interconnected, both between upper and lower parts and between neighbouring folds, and gradually grow into a fully evolved "curtain" structure. The largest "curtain folds" are the places from which flux later pinches off and migrates into the sample volume in the form of flux tubes, to create a multiply-connected topology of the IS superconducting phase.

To qualitatively describe the flux structures observed in Fig. 1, two basic MS are initially sufficient: catenoid and Scherk's (1st). As evident from Fig. 2, a set of Scherk's surfaces resembles an edge curtain structure, while a catenoid is the basic form of a flux bundle within the sample volume. These surfaces are basic, nontrivial surfaces of revolution and translation.

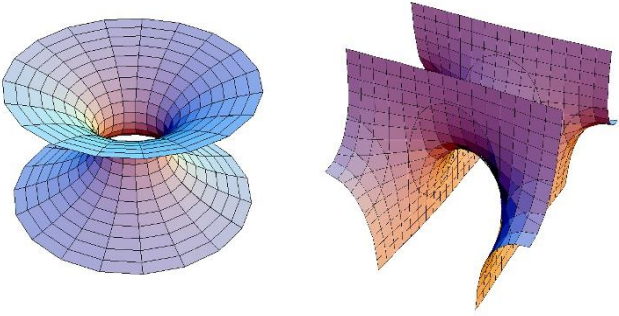

FIG. 2. Catenoid and 1st Scherk's MS.

Although other appropriate MS types, like Scherk's 2nd surface or others from associate family of MS, could be realised in more complex forms of type-I superconductors IS, we restrict our analysis to mentioned the simplest ones (catenoid and Scherk's 1st).

Assuming $z$ axis oriented along the sample thickness, with $z = 0$ at the sample equator (middle), these surfaces can be expressed in analytical form as

$$\cosh\left(\frac{z}{R}\right) = \frac{\sqrt{x^2 + y^2}}{R} \qquad (6)$$

$$\exp\left(\frac{y}{L}\right) = \frac{\cos(x/L)}{\cos(z/L)}, \qquad (7)$$

where $R$ is the central radius of the catenoid, and $L$ is the characteristic length scale of the Scherk's surface.

### A. Catenoid

The amount of the magnetic flux carried by each individual flux tube in a type-I superconductor is limited only by the criterion of being a multiple of the elementary flux quanta $N\Phi_0$. Assuming a catenoid geometry for the tube, the amount of contained flux is then

$$\Phi_{catenoid} = \pi R^2 H_c = N\Phi_0 . \qquad (8)$$

The free energy fraction associated with catenoid like tube geometry is proportional to its surface area and volume as

$$\Delta F_{catenoid} \approx \left(\delta S_{catenoid} + \pi R^2 d\right)\frac{H_c^2}{8\pi} . \qquad (9)$$

Analytical solution based on eq. (6) yields for $S_{catenoid}$

$$S_{catenoid} = \pi R^2 \left[\frac{d}{R} + \sinh\left(\frac{d}{R}\right)\right] . \qquad (10)$$

After substituting Eq. (10) into Eq. (9), we find that for each particular sample thickness ($d$) there is a region of "lowest energy" $d/R$ tube ratios, with an absolute minimum depending on the $\delta/d \approx \xi/d$ ratio. This variation with $R/d$ is shown in Fig. 3 for various $\delta/d$.

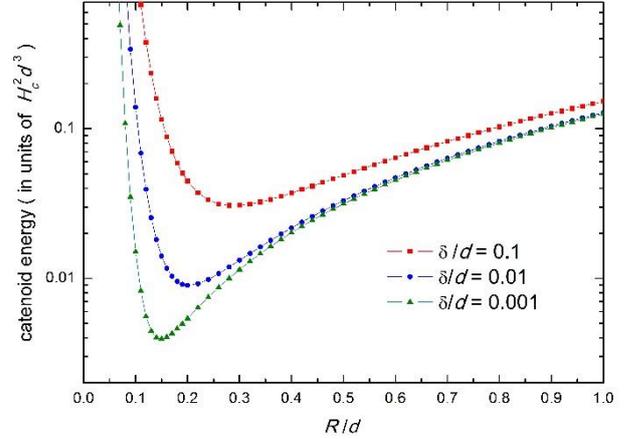

FIG. 3. Catenoid like flux tube energy for three different $\delta$ / thickness ratios.

This result indicates, that for type-I superconductors planar samples, with given thickness ($d$) and material parameter ($\delta$), optimal size ($R/d$)$_{min}$ of catenoid like flux tubes exists, and that smaller or larger tubes are energetically less favourable (see Table I.).

TABLE I. Superconductive parameters and calculated characteristic $R/d$ for selected type-I superconductors and samples thicknesses.

|  | Pb | Sn | In | Al |
|---|---|---|---|---|
| $\xi_0$ (μm) | 0.083 | 0.230 | 0.360 | 1.600 |
| $\lambda_0$ (μm) | 0.035 | 0.034 | 0.065 | 0.016 |
| $d$ (μm) | 10 | | | |
| $\delta/d$ | 0.0048 | 0.0196 | 0.0295 | 0.1584 |
| ($R/d$)$_{min}$ | 0.18 | 0.22 | 0.24 | 0.31 |
| $d$ (μm) | 20 | | | |
| $\delta/d$ | 0.0024 | 0.0098 | 0.0147 | 0.0792 |
| ($R/d$)$_{min}$ | 0.17 | 0.20 | 0.21 | 0.28 |
| $d$ (μm) | 50 | | | |
| $\delta/d$ | 0.00096 | 0.00392 | 0.00590 | 0.03168 |
| ($R/d$)$_{min}$ | 0.15 | 0.18 | 0.19 | 0.24 |

Combining optimal $R/d$ parameter with critical field strength expected at sample middle plane, we estimated typical number ($N_{min}$) of elementary magnetic flux quanta associated with individual catenoid like flux tubes. To cross-check consistency of used superconductive parameters, we also calculated theoretical number of flux quanta ($N_{GL}$) for the flux tubes of the same size directly from coherence and penetration lengths, utilizing their relation to the critical field and elementary flux quanta within Ginzburg-Landau theory (see Table II.).

TABLE II. Estimated number of elementary magnetic flux quanta associated with energetically most favorable catenoid like flux tubes for selected type-I superconductors and samples thicknesses.

|  | Pb | Sn | In | Al |
|---|---|---|---|---|
| $\mu_0 H_{c0}$ (T) | 0.080 | 0.031 | 0.023 | 0.011 |
| $d$ (μm) | 10 | | | |
| $N_{min}$ | 393 | 227 | 201 | 160 |
| $N_{GL}$ | 394 | 219 | 87 | 132 |
| $d$ (μm) | 20 | | | |
| $N_{min}$ | 1404 | 753 | 616 | 524 |
| $N_{GL}$ | 1407 | 723 | 266 | 433 |
| $d$ (μm) | 50 | | | |
| $N_{min}$ | 6836 | 3814 | 3153 | 2406 |
| $N_{GL}$ | 6846 | 3662 | 1364 | 1989 |

As indicated by $N_{min}$ and $N_{GL}$ values, superconductive parameters values commonly listed in literature are not mutually fully consistent, particularly for indium, but discrepancies are acceptable for further analysis.

The $(R/d)_{min}$ ratio isn't the only notable one. To preserve genuine catenoid form, local field strength anywhere within flux tube must remain higher than $H_{c2}$ to prohibit spontaneous superconductivity nucleation close to samples surfaces ($z = \pm d/2$). Assuming thermodynamic field strength ($H_c$) inside flux bundle at sample middle plane ($z = 0$), then from flux conservation and Eq. (6) we find condition

$$\cosh^2\left(\frac{d}{2R}\right) = \frac{1}{\sqrt{2}}\frac{1}{\kappa}, \qquad (11)$$

where $\kappa = \lambda/\xi$ is Ginzburg–Landau (G-L) parameter. Eq. 11 defines low limit $(R/d)_{cat}$ which is sample thickness independent. Contrary, the absolute upper limit on the $R/d$ ratio is defined by a global free energy minimization and associated periodicity of IS [3,4]. Comparing IS periodicity, given as

$$P = \sqrt{\frac{\delta d}{f_L(h)}}, \qquad (12)$$

where $f_L(h)$ is Landau function of reduced field strength ($h$) [12], with catenoid diameter at sample surface, i.e. combining Eq. (12) and Eq. (6), we obtain inequality

$$\sqrt{\frac{\delta d}{f_L(h)}} \geq 2R\cosh\left(\frac{d}{2R}\right), \qquad (13)$$

where periodicity of IS is assumed to be at least twice larger than catenoid surface radius. For further analysis it is practical to reformulate this inequality and solve limit situation in the form

$$\sqrt{\frac{\delta}{d}\frac{1}{f_L}} = \frac{2R}{d}\cosh\left(\frac{d}{2R}\right). \qquad (14)$$

Solution of Eq. (14) define the maximal ($R/d$) for which catenoid like flux tubes could exist wholly up to sample surfaces. Exploiting $R/d$ as a free parameter, the right-side value of Eq. (14) reaches minimum at $R/d = 0.42$ and fulfilment of inequality is predominantly controlled by a left-side value. Utilizing Landau function field dependence in the form [13]

$$f_L(h) = \frac{1}{4\pi}\Big[(1+h)^4\ln(1+h)+(1-h)^4\ln(1-h)- \\ -(1+h^2)^2\ln(1+h^2)-4h^2\ln(8h)\Big], \qquad (15)$$

and parameters of superconductive materials listed in Table I., we can identify two field strength regions, where inequality is satisfied. The first one, in vicinity of low $H_a$ and second close to critical field intensity. While in both regions inequality Eq.(13) is formally satisfied, only first region is relevant for purpose of this work, since in the second one it is the normal phase fraction which is the source of large observed periodicity and no catenoid like normal zones could exist there. The maximal field strength at which catenoid like flux bundles could exist, labelled as ($h_{i\_max}$), found by numerical evaluation of Eq.(14) left side are listed in Table III.

TABLE III. Limits for possible catenoid like flux bundles existence for selected superconductors and thicknesses.

|  | Pb | Sn | In | Al |
|---|---|---|---|---|
| $(R/d)_{cat}$ | 0.67 | 0.35 | 0.38 | 0.18 |
| $d$(μm) | 10 | | | |
| $(h)_{i\_max}$ | 0.053 | 0.138 | 0.193 | <1 |
| $d$(μm) | 20 | | | |
| $(h)_{i\_max}$ | 0.083 | 0.256 | <1 | <1 |
| $d$(μm) | 50 | | | |
| $(h)_{i\_max}$ | 0.162 | <1 | <1 | <1 |

Combining listed $R/d$ limits and applied fields we can deduce the form of expected flux (normal zones) patterns observed on the samples surfaces in magneto-optical experiments. For example, the lead samples could attain pure catenoid form only for $R/d > 0.67$, i.e. for all evaluated thicknesses the energetically favourable bundle size would typically flare-out near sample surfaces, creating isolated lattice of macroscopic flux structures with corrugated pattern on perimeter and superconducting islands inside them (see Fig. a,b,c, page 23 in [2]). For other evaluated superconductors $R/d$'s discrepancies between energetically ideal bundles vs. genuine catenoids are not that pronounced. Considering fairly slow grow of catenoid energy, particularly for thin (10 μm) samples, pure form of catenoid like flux bundles could exist (like in [11]).

However, this doesn't necessitate that such structure must be observed in all experiments. The amount of contained flux (size of normal zones) is dominantly driven by dynamics of flux penetration into the sample. For example, the flux penetrating from samples edges, through region of superconducting shielding currents (i.e. so-called geometrical barrier), into fully superconducting samples ($T \ll T_c$) is controlled mainly by samples edge properties, both superconducting and geometrical, while IS nucleated in the course of $H_a$ decrease from values above critical, would be determined mainly by sample superconducting properties and purity [14]. In addition, when forming IS via cooling sample through $T_c$ in weak external magnetic field, resulting IS structures would depend predominantly on the rate of temperature decrease. In all considered cases strictly perpendicular direction of applied magnetic field is considered, since any significant deviation from such orientation would cause reduction of flux bundles symmetry and general tendency to form strip like IS structures [13].

### B. Scherk's

The most common magneto-optical experiments set-up comprises the study of IS established via continuous $H_a$ increase acting upon initially fully superconducting samples ($T \ll T_c$), with magnetic flux penetrating into bulk from samples edges. In such scenario the local free energy balance between individual Scherk's MS, forming edge folds, and detached catenoid like flux tubes could be considered to describe basic dynamics of such process.

Let us assume the edge curtain structure to be a periodic set of an identical Scherk's MS expressed in analytical form by Eq. 7, where $x$ axis is parallel with sample edge, $y$ axis is pointing into the sample and origin of coordinate system is shifting in $y$ direction throughout $H_a$ increase. The direct consequence of this assumption is a lower limit on the curtain fold characteristic length scale and periodicity of $L > d/\pi$, which also indicates that ultimately all IS structures scales are linked to sample thickness. For $H_a$ well below $H_p$, the fold periodicity is significantly larger and each fold is only partially evolved, i.e. it may only partially cut through the sample volume near upper and lower sample planes, and individual partial curtain folds are not interconnected. At this stage, Scherk's central point (i.e. origin of coordinate system in Eq. 7 representation) is situated outside the sample and clearly no magnetic flux could penetrate into the sample.

Near $H_p$, $L$ decreases towards $d/\pi$ and the folds are almost fully evolved. In this situation, the individual folds are already interconnected near the sample's upper and lower surfaces. At the sample equator ($z = 0$), however, each of the folds is still strictly individual, while Scherk's central point shifted into the sample volume.

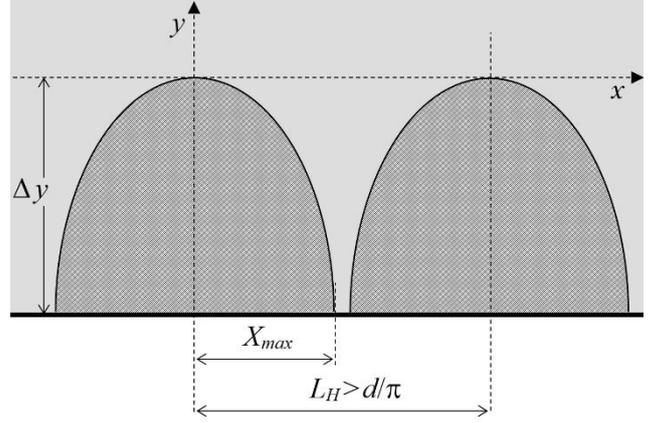

FIG. 4. Sketch of two, flux conveying, Scherk's MS equatorial plane ($z = 0$) cuts, for $H_a$ approaching $H_P$.

The distance $\Delta y$ how far is central point shifted into the sample volume is linked with $X_{max}$ and $L_H$ as

$$\Delta y = (-L_H) \log\left[\cos\left(\frac{X_{max}}{L_H}\right)\right], \quad (16)$$

where $L_H$ is instantaneous, i.e. applied field dependent fold periodicity (or Sherks's scale parameter) and $X_{max}$ is distance between Scherk's fold cuts through the sample edge and fold symmetry axis at $z = 0$, which is always smaller than $L_H/2$ (see Fig. 4).

To calculate essential parameters of curtain like edge structure, as the amount of magnetic flux carried by each fold and energy balance of S-N interface, we define two dimensionless parameters: $\varepsilon = z/L_H$ and $\eta = x/L_H$. They attain values from intervals $\varepsilon \in (-d/(2L_H), d/(2L_H))$ and $\eta \in (-X_{max}/L_H, X_{max}/L_H)$. For fully evolved Scherk's folds ($X_{max} \to \pi L_H/2$) and critical periodicity ($L_H \to d/\pi$), the surface area becomes infinite, i.e. S-N interface energy diverge.

The amount of flux carried by each fold is proportional to the equatorial Scherk's surface cut ($\varepsilon = 0$), which can be expressed as

$$S_{flux} = 2L_H^2 \int_0^{\eta_{max}} \eta \, \mathrm{tg}\, \eta \, d\eta, \quad (17)$$

where and $\eta_{max} = X_{max}/L_H$. To calculate S-N interface area of Scherk's surface, it is useful to parametrize it as

$$x = L \arccos(u) \,;\, y = L \log\left(\frac{u}{v}\right) \,;\, z = L \arccos(v), \quad (18)$$

finding

$$S_{S-N} = L^2 \iint \sqrt{\frac{u^2 - u^2 v^2 + v^2}{(1-u^2)(1-v^2)}} \, \frac{du}{u} \frac{dv}{v}. \quad (19)$$

Reverting to $\varepsilon$ and $\eta$ parameters and field dependent $L_H$ we find

$$S_{S-N} = L_H^2 \iint \sqrt{1 + tg^2 \varepsilon + tg^2 \eta}\, d\varepsilon\, d\eta \quad . \tag{20}$$

The free energy fraction associated with Scherk's like edge structure is proportional to its surface area and volume as

$$\Delta F_{Scherk} \approx \left( \delta S_{S-N} + d\, S_{flux} \right) \frac{H_c^2}{8\pi} \quad . \tag{21}$$

Even though solution of Eq. 17 could be expressed via even Bernoulli numbers, both Eq. 17 and 20 have been evaluated numerically. To select integration boundaries in Eq. 20 we exploited mathematical symmetry of Scherk's surface reflected correspondingly in the integrand. In the course of Scherk's like edge structure evolution, seeking to minimize S-N interface area, the central point ($\Delta y$) and scale parameter ($L_H$) are self-determined parameters. For that reason, both $\varepsilon$ and $\eta$ parameters are equally significant and $\varepsilon_{max} \approx \eta_{max}$ should be assumed in Eq. 20 integration. Variation of energy associated with each Scherk's fold vs. $\eta_{max}$, i.e. folds evolution stage, is shown in Fig. 5 for typical $\delta/d$.

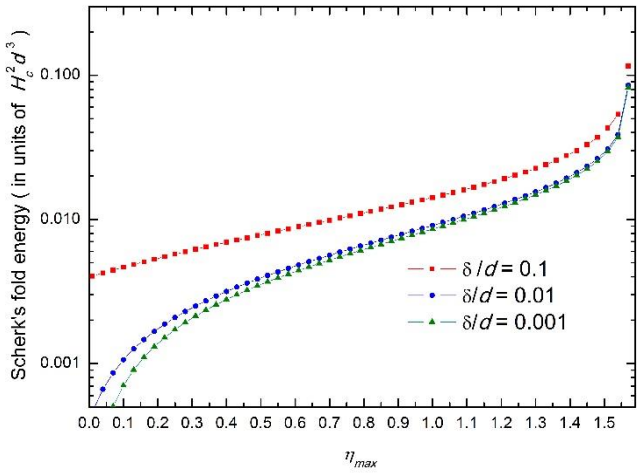

FIG. 5. Scherk's like flux structure energy for three different $\delta$ / thickness ratios.

Comparing energies associated with Scherk's fold and ideal catenoid flux tube we observe that there is always a possibility to separate flux tubes from edge folds. Table IV lists lower limits of $\eta_{max}$ (or $\varepsilon_{max}$) at which such process is energetically allowed for selected superconductors and thicknesses.

TABLE IV. Lower limits for $\eta_{max}$ to energetically permit disentanglement of optimal catenoid like flux bundles from Scherk's like edge folds. Ideal $(R/d)_{min}$ and superconductive parameters are listed in TABLE I.

|  | Pb | Sn | In | Al |
|---|---|---|---|---|
| $d$ (μm) | 10 | | | |
| $\delta/d$ | 0.0048 | 0.0196 | 0.0295 | 0.1584 |
| $\eta_{max}$ | 0.83 | 1.15 | 1.24 | 1.47 |
| $d$ (μm) | 20 | | | |
| $\delta/d$ | 0.0024 | 0.0098 | 0.01475 | 0.0792 |
| $\eta_{max}$ | 0.70 | 0.99 | 1.09 | 1.40 |
| $d$ (μm) | 50 | | | |
| $\delta/d$ | 0.00096 | 0.00392 | 0.00590 | 0.03168 |
| $\eta_{max}$ | 0.55 | 0.79 | 0.88 | 1.25 |

While energy criterion is mandatory, the flux conservation in the process should be also considered. For well evolved edge flux folds each catenoid like bundle separates from individual folds, which must contain at least equal or larger amount of magnetic flux. Since both structures equatorial field strength is critical, flux conservation criteria reduces to $S_{flux} > \pi R^2_{min}$, where $S_{flux}$ is defined by Eq. 17 and $R_{min}$ is central radius of ideal catenoid like flux bundle. Table V allows comparison of Scherk's like edge folds $S_{flux}$, evolved up to $\eta_{max}$, i.e. when energy criterion became satisfied, with ideal catenoid bundle equatorial cross-section ($\pi R^2_{min}$).

TABLE V. Comparison of flux carrying cross sections, with $H_c$ magnetic field strength, of Scherk's like edge fold at $\eta_{max}$ listed in TABLE IV., vs. energetically most favorable catenoid like flux tubes.

|  | Pb | Sn | In | Al |
|---|---|---|---|---|
| $d$ (μm) | 10 | | | |
| $S_{flux}$ (μm²) | 16.2 | 27.6 | 32.7 | 60.6 |
| $\pi R^2_{min}$ (μm²) | 10.2 | 15.2 | 18.1 | 30.2 |
| $d$ (μm) | 20 | | | |
| $S_{flux}$ (μm²) | 52.0 | 84.2 | 99.2 | 190.3 |
| $\pi R^2_{min}$ (μm²) | 36.3 | 50.3 | 55.4 | 98.5 |
| $d$ (μm) | 50 | | | |
| $S_{flux}$ (μm²) | 244.4 | 379.4 | 440.2 | 833.0 |
| $\pi R^2_{min}$ (μm²) | 176.7 | 254.5 | 283.5 | 452.4 |

As exhibited by calculated values in Table V, for all considered superconductors and sample thicknesses, the flux criterion is always satisfied when energy criterion is met. This proves possibility to detach catenoid like flux bundle from strictly individual Scherk's like edge folds in process, closely analogous to soap film bubble formation by separating from a cylindrical soap film, when a critical length-to-radius ratio (typically 3.1) is reached [15].

### C. Numerical estimate of $H_p$

The majority of experiments probing magnetic flux penetration into the Type-I superconductors addresses the value of the first penetration field $H_p$. This field is obviously strongly dependent on the sample geometric demagnetization factor, i.e. sample size and actual geometry. In this work we do not concentrate on determining precise $H_p$ field value, although from constructed MS model we are able to estimate $H_p$ for rectangular strip samples with large length vs. width ($L_x/L_y$) ratios.

As we established in preceding paragraphs, the flux bundles propagating into sample volume separate in fact from individual edge folds. For that reason, the width of evolved Scherk's fold could be regarded as a basic sample entity (cell). Necessitating the flux conservation within this cell, the following equation must be satisfied

$$2S_{flux}H_c = L_H L_y H_p \quad . \qquad (22)$$

The reduced penetration field ($h_p$) in this model is therefore primarily inversely proportional to sample width ($1/L_y$) with nontrivial $S_{flux}/L_H$ modulation. From described MS model, the $S_{flux}/L_H$ parameter, including relation to sample thickness, could be worked out. Using $L_H$ defined through $\eta_{max}$ and $S_{flux}$ at the critical state, i.e. when flux penetration became energetically feasible (Table IV), the estimate of $h_p$ is straightforward. Table VI. lists calculated $h_p$ for selected superconductors and $L_y/d$ ratios.

TABLE VI. Estimated reduced penetration field values for selected superconductors and $L_y/d$ ratios.

|  | Pb | Sn | In | Al |
|---|---|---|---|---|
| $L_y/d$ | 10 | | | |
| $h_p$ | 0.054 | 0.127 | 0.162 | 0.356 |
| $L_y/d$ | 20 | | | |
| $h_p$ | 0.018 | 0.042 | 0.054 | 0.133 |
| $L_y/d$ | 50 | | | |
| $h_p$ | 0.004 | 0.010 | 0.012 | 0.033 |

## IV. DISCUSSION

Introduced MS model is founded upon plausible idea, that energy minimization is equally important both at local and global samples scales. For comparison of modelled parameters with experimental observations several restrictions have to be observed. At first, MS model assumes ideally pure type-I superconductor samples, both in chemical and metallurgical meanings, in conjunction with their strict rectangular geometrical shape. Secondly, the high field homogeneity and orientation perpendicular to sample surface is assumed, since even miniscule field orientation misalignment would prefer strip like flux structures grow [13]. Additionally, fairly slow applied field ramp-up is required, as too fast field ramp would disallow evolution of internal individual flux structures into semi equilibrium state at instantaneous field strength. Finally, only samples without previous field history, i.e. cooled bellow $T_c$ in magnetic field free environment, are of relevance, in conjecture with complete absence of any magnetic flux within samples, either permanently pinned to local defects or trapped during sample cooling. Evident example of field history and orientation impact upon observed IS flux structures, while examining identical sample, is shown in Ref. [16]. Although no realistic experimental set-up could fully met listed requirements, the basic comparisons with experimental results are possible.

Only few experiments, aiming on first penetration field studies, used rectangular strip samples and are therefore suitable to test our MS model quantitatively. They were usually performed by direct induced voltage signal measurement, originating from magnetic flux penetration through sensitive element (coupling coil or loop) [17]. For tin samples with $L_y/d$=11.1 reduced penetration field $h_p$=0.17 have been observed at 1.7 K, which is in good agreement with values listed in Table VI. However, analogous earlier tin experiments performed at higher temperatures (1.9 to 3.5 K) with $L_y/d$=29.3 samples, yield significantly higher reduced penetration fields, ranging between 0.19÷0.34 [18]. Substantial difference of observed $h_p$ at temperatures close to $T_c$ should be attributed to inapplicability of adopted δ/d parameter, and corresponding $\eta_{max}$ values listed in Table IV, when critical field $H_c(T)$ and coherence length ξ(T) scaling adopting G-L theory have been assumed only. Our own experiments, using strip samples cut from 50 μm thick tin sheets with $L_y/d$=16, measured at 0.3 K temperature, lead to $h_p$=0.18÷0.23, i.e. values in better agreement with Table VI values. Differences observed in our experiments should be attributed to poor control of samples edge rectangularity.

Majority of magneto-optic experiments exploited disk shape samples, therefore only qualitative comparisons of flux structures shapes and geometries with MS model are conceivable. Furthermore, each magneto-optic measuring system is set to be sensitive at particular magnetic field strength, i.e. measured flux pattern dimensions should be referred to this setting. As demonstrated in Ref. [5], the measured flux tube diameter could vary by a factor 2÷3 times when measured at low or high field strength sensitivity limit and observed flux structure shapes represent in reality cross section of actual flux bundle geometry at set field sensitivity, not necessarily at actual samples surface. Nevertheless, taking into consideration uncertainty of mercury coherence length ($\xi_0 \approx 130$ nm) at 0.41 reduced temperature, the observations in Ref [11] (TABLE I and Fig. 8 therein), particularly for thin samples (47 and 106 μm) and low reduced applied field, are in fair agreement with MS catenoid shape, and observed flux bundles diameters differ from ideal catenoid prediction by less than factor of two. In real magneto-optic experiments, the considerable variation of individual bundles linear scales (diameters) are being typically observed. This variation arises naturally from the need to apply significantly higher magnetic field, well above first penetration field, to observe both edge and internal flux regions of tested samples. At those conditions direct comparison with presented MS model is not viable.

Qualitatively presented MS model plausibly explains complicated "flower" like patterns observed in thick (0.68 mm) and large lead disk (diameter 12 mm) samples reported in [2]. Particularly Fig. 2.9 a,b (page 23) therein, are typical examples. When the first penetration field is reached in alike samples, they thickness force a very large flux bundles to penetrate from edges. After those bundles migrate into sample interior (central region), while remaining spatially well separated, they display approximate axial symmetry with curtain-like edges. This could be understood as a consequence of local energy minimization for each individual bundle. Axial symmetry originates from catenoid like structure established deep inside sample volume (unobservable), while bundle flare out near sample surfaces combine them with partially developed Scherk's folds.

V. CONCLUSIONS

We have approached the problem of the origin of the IS flux structures in type-I superconductors shapes and sizes from the point of view of MS properties as a consequence of local energy minimization, as an extension to a global minimal energy approach. Among these, a catenoid structure is seen to represent an idealized flux bundle, and Scherk's surface the IS structure at the sample edge.

A cursory comparison indicates the possibility for further development of this approach towards understanding the origin of the more complicated IS structures, not only in Type-I superconductors, but possibly also in other physical systems where the condensate phase, formed after phase transition, coexists with normal phase.

**The support** of Civic association "My sme niekto iní" is acknowledged, as also that of the Foundation for Science & Technology of Portugal (POCTI/FNU/10033/2001, POCTI/FIS/39067/2001).